\documentclass[aps,prb,twocolumn,floatfix,showpacs,citeautoscript,superscriptaddress,longbibliography,hyperlinks]{revtex4-2}
\usepackage{amsmath}
\usepackage{amssymb}
\usepackage{amsfonts}
\usepackage[colorlinks=true,citecolor=blue]{hyperref}
\usepackage{graphicx}
\usepackage{enumitem}
\setlist[enumerate]{leftmargin=6mm}
\usepackage[caption = false]{subfig}
\usepackage{braket}
\usepackage{siunitx}
\usepackage{tikz}
\usepackage{bbm}
\usepackage{circuitikz}
\usepackage[capitalise]{cleveref}
\usetikzlibrary{%
shapes.geometric,calc,intersections,%
decorations.markings,decorations.text,decorations,%
patterns,decorations.pathmorphing,fpu,%
decorations.pathreplacing,%
positioning,automata,shapes.multipart,circuits,%
circuits.ee, circuits.ee.IEC, shapes.gates.ee,%
arrows.meta,arrows,%
fadings
}

\newcommand{\mytodo}[1]%
{{\todo[inline,backgroundcolor=blue!10!white]{#1}
}}
\newcommand{\me}{\mathrm{e}}
\newcommand{\mi}{\mathrm{i}}
\newcommand{\md}{\mathrm{d}}

\DeclareMathOperator{\sinc}{sinc}

\usepackage[normalem]{ulem}

\begin{document}

\title{Wigner representation of Andreev-reflected charge pulses}

\author{Benjamin Roussel}
\affiliation{Department of Applied Physics,
	Aalto University, 00076 Aalto, Finland}
\author{Pablo Burset}
\affiliation{Department of Theoretical Condensed Matter Physics\char`,~Universidad Aut\'onoma de Madrid, 28049 Madrid, Spain}
\affiliation{Condensed Matter Physics Center (IFIMAC), Universidad Aut\'onoma de Madrid, 28049 Madrid, Spain}
\affiliation{Instituto Nicol\'as Cabrera, Universidad Aut\'onoma de Madrid, 28049 Madrid, Spain}
\author{Christian Flindt}
\affiliation{Department of Applied Physics,
	Aalto University, 00076 Aalto, Finland}

\begin{abstract}
Recent experiments have shown that the edge states of a quantum Hall sample can be coupled to a superconductor, so that incoming electrons in the edge states can be Andreev converted by the superconductor as coherent superpositions of an electron and a hole. In parallel, single-electron emitters that operate in the gigahertz regime have been realized on quantum Hall edge states. Motivated by this remarkable experimental progress, we here analyze the Andreev reflections of incoming charge pulses on the interface with a superconductor. To this end, we employ a Wigner function representation that allows us to visualize the response function of the interface both in the time and in the frequency domain. We analyze the response of an interface between a normal-metal and a singlet or triplet superconductor with and without an insulating barrier in between them. As a special case, we analyze the Andreev reflections of clean single-particle excitations that are generated by the application of lorentzian-shaped voltage pulses to the contacts of the inputs. Our predictions may be tested in future experiments with edge states coupled to superconductors.
\end{abstract}

\maketitle

\section{Introduction}
Electron quantum optics is an emerging field that applies concepts and techniques from quantum optics to
investigate the quantum properties and interference of individual charges in mesoscopic conductors~\cite{bocquillon2014electron,Splettstoesser:2017,Roussel_2017,Bauerle:2018,Edlbauer2022}.
Clean single-electron excitations can now be generated and emitted into
a mesoscopic conductor without accompanying electron-hole
pairs~\cite{feve2007demand,dubois2013minimal,jullien2014quantum,Assouline:2023}.
These dynamic excitations can be probed at the single-electron level in
tomographic
experiments~\cite{jullien2014quantum,bisognin2019quantum,Assouline:2023,fletcher2019continuous},
and they can be made to interfere
~\cite{bocquillon2013coherence,dubois2013minimal} and
interact~\cite{Ubbelohde2023,Fletcher:2023,Wang:2023} at quantum point
contacts that function as electronic beam splitters. In addition to
fundamental investigations of charge pulses in mesoscopic conductors,
electron quantum optics also holds promises for applications such as for
current standards~\cite{giblin2012towards} and electronic flying qubits~\cite{Aluffi:2023}.

Up until now, experiments in electron quantum optics have been carried out with electrons and holes in metals or semiconductors~\cite{feve2007demand,bocquillon2013coherence,dubois2013minimal,jullien2014quantum,Assouline:2023,bisognin2019quantum,fletcher2019continuous,Ubbelohde2023,Fletcher:2023,Wang:2023}.
However, the addition of superconductors would enable additional types
of processes, for example, by allowing for Andreev reflections whereby
an electron is converted coherently into a hole~\cite{acciai2019levitons,Sassetti_2020,aBurset_short,Burset_long}. Such a conversion process of a particle into its anti-particle has no counterpart in quantum optics with photons and would thus be unique to electron quantum optics~\cite{Vanevic_2015,averin2020time,Martin_2022,bertin2023current}.
The inclusion of superconductivity to electron quantum optics seems promising, and it is motivated by several recent experiments where the edge states of a quantum Hall sample have been coupled to a superconductor~\cite{Zulicke_2005,Ustinov_2007,Schonenberger_2012,Rokhinson_2015,Calado2015,BenShalom2016,Amet_2016,Lee_2017,Das_2018,Finkelstein_2020,Das_2021,Shabani_2022,vignaud2023evidence,Zhao:2023}.
Combined with techniques from electron quantum optics, one might explore the dynamic processes at the interfaces between normal metals and superconductors in the high-frequency regime~\cite{Clarke_2014,Nazarov_2017,Nazarov_2019,zhang2019perfectcar,Akhmerov_2022,Klinovaja_2022,kurilovich2022disorder,Schmidt_2023,Oreg_2023,Balseiro_2023}.
Such a setup would enable quantum superpositions of states with different numbers of charges~\cite{Dasenbrook:2015}. For example,
Andreev processes would allow for the transformation of an electron into a
superposition of an electron and a hole, which could find use for
interferometric measurements and sensing~\cite{sequoia2022}.

\begin{figure}
	\input{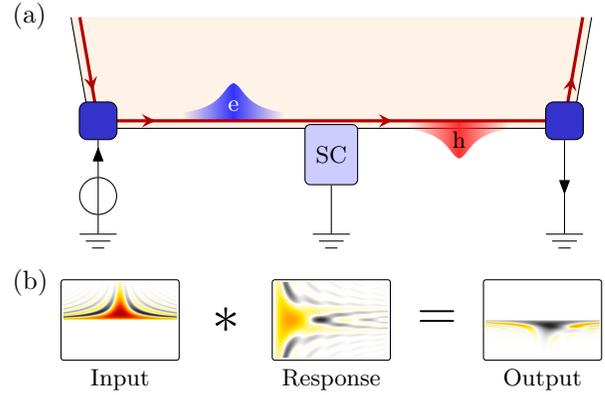}
	\caption{Wigner representation of Andreev conversions.
		(a)~Single-electron pulses are injected into a quantum Hall edge state and are scattered off the interface with a superconductor. The charge pulses can be either normal-reflected or Andreev converted at the interface, or they can be transmitted into the superconductor above the gap.
		(b) The response of the interface can be represented by a Wigner
		function describing the conversion of an input state to an
		output state, both represented by their Wigner function.
	}
	\label{fig/abstract}
\end{figure}

Figure~\ref{fig/abstract} illustrates the essential idea of our work. We
consider voltage pulses that are applied to the edge states of a quantum
Hall sample, so that single-electron excitations are emitted towards the
interface with a superconductor. 
Here, they are either Andreev converted or normal-reflected back into
the edge states, or they are transmitted into the superconductor above
the gap. The wave functions of the incoming and the outgoing particles
can be visualized by their Wigner functions in the mixed time and
frequency
representation~\cite{ferraro2013wigner,Ferraro:2014,Roussel:2021}, and
they are related by the response function of the superconducting
interface as shown in the figure.

The rest of our article is organized as follows. In
Sec.~\ref{sec/1dint}, we employ a circuit model to describe the
interface between a metal and a superconductor in one dimension. In
Sec.~\ref{sec/coherence}, we define the single-electron coherence and
its Wigner representation. In Sec.~\ref{sec/respfunc}, we analyse the response
functions for singlet and triplet superconductors with and without an
insulating barrier. In Sec.~\ref{sec/leviton}, we consider
lorentzian-shaped voltage pulses that generate clean single-particle
excitations known as levitons. We then investigate how levitons are
Andreev converted or normal-reflected on the superconductor. Finally, in
Sec.~\ref{sec/conclusion}, we present our conclusions together with an
outlook on possible avenues for further developments. Some technical
details of our calculations are
deferred~to~App.~\ref{app/deriveq/respNS}. 

\section{Superconductor interface}
\label{sec/1dint}

\subsection{Hamiltonian}

In this section, we introduce a one-dimensional model of an interface between a normal-metal and a superconductor~\cite{blonder1982transition}. We consider conventional $s$-wave (or singlet)
superconductors as well as unconventional $p$-wave (or triplet) superconductors that are connected to a normal-metal either directly (an NS junction) or with an insulator between them (an NIS junction). We then derive a circuit description of the superconductor, which we can solve for arbitrary incoming electronic wave functions. 

We start by deriving a low-energy approximation of the BCS
Hamiltonian for excitations that are generated close to the Fermi level. To this end, we consider the interacting part of the BCS Hamiltonian in one dimension,
\begin{equation}
	\hat H_{\text{BCS}} =
	\int \frac{\md k}{2 \pi}
	\left(
		\Delta_{\sigma'\sigma}^* \hat c_{-k,\sigma} \hat c_{k,\sigma'}
		+
		\Delta_{\sigma\sigma'} \hat c^\dagger_{k,\sigma} \hat c^\dagger_{-k,\sigma'}
	\right),
	\label{eq/hamiltonian/bcs/effective}
\end{equation}
with $\Delta_{\sigma\sigma'}$ being the pair potential and $\hat c^\dagger_{k,\sigma}$ and $\hat c_{k,\sigma}$ are the usual fermionic operators for electrons with spin $\sigma$. For simplicity, we will consider $s$-wave pairing. In this case, $\Delta$ is an even function of the momentum~\cite{Sigrist_RMP}. Furthermore, we will also make the simplifying assumption that $\Delta$ does not depend on the momentum. 

\begin{figure}
	\begin{tikzpicture}
	\def\xmax{1.7}
	\def\ymax{2.3}

	\def\xfermi{1.1}
	
	\pgfmathsetmacro{\paraa}{\ymax/\xmax/\xmax}
	\pgfmathsetmacro{\yfermi}{\paraa*\xfermi*\xfermi}
	\pgfmathsetmacro{\derivfermi}{2*\paraa*\xfermi}

	\foreach \xshift/\spintext/\spin/\colspin in {0/U/\uparrow/green, 4.5/D/\downarrow/red} {
		\begin{scope}[shift={(\xshift, 0)}]
			\draw[->] (-\xmax, 0) -- (\xmax, 0)
				node[below] {$k_{\spin}$};
			\draw[->] (0, {-0.1*\ymax}) -- (0, \ymax)
				node[right] {$E_{\spin}$};

			\draw (-\xmax, \ymax) parabola[parabola height=-\ymax cm] (\xmax, \ymax);

			\fill[\colspin,opacity=0.2] (-\xfermi, \yfermi)
				parabola[parabola height=-\yfermi cm]
				(\xfermi, \yfermi);
			\foreach \dir/\dirtext in {-1/L, 1/R} {
				\draw[\colspin!60!black, thick] ({\dir*\xfermi-\dir*\yfermi/\derivfermi}, 0) --
					({\dir*\xmax}, {\yfermi + (\xmax-\xfermi)*\derivfermi});
				\coordinate (fermi-\dirtext-\spintext) at ({\dir*\xfermi}, \yfermi);
			}
		\end{scope}
	}

	\draw[stealth-stealth, thick, opacity=0.2, blue] ($(fermi-L-U) + (0.1, 0.1)$)
		to[bend left=20] ($(fermi-R-D) + (-0.1, 0.1)$);
	\draw[stealth-stealth, thick, opacity=0.2, blue] ($(fermi-R-U) + (0.1, 0)$)
		to[bend left] ($(fermi-L-D) + (-0.1, 0)$);
\end{tikzpicture}
	\caption{Low-energy description.
		Close to the Fermi
		level, the dispersion relation can be linearized and left- and
		right-movers are coupled by the superconductor. For 
		singlet superconductors, as shown, opposite spins are coupled.
	}
	\label{fig/dispersion}
\end{figure}
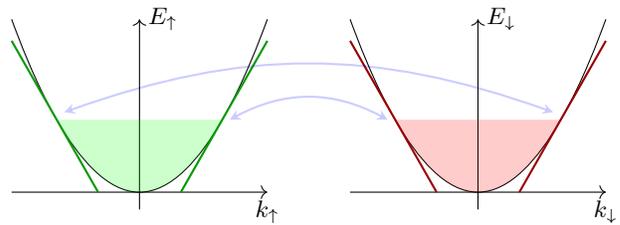

In the following, a real-space representation of the Hamiltonian will be
convenient since we consider the interface between a superconductor and a normal-metal. We focus on low-energy excitations near the
Fermi level and can therefore treat left and right movers separately, since they form two
independent chiral and linear branches as illustrated in Fig.~\ref{fig/dispersion}. We then define the corresponding field operators as
\begin{equation}
	\hat\psi_{\sigma,L}(x)
	=
	\int_{0}^{\infty} \md k\, \hat c_{\sigma,k} \, \me^{\mi (k-k_F) x} 
 \end{equation}
 and
 \begin{equation}
	\hat\psi_{\sigma,R}(x)
=
	\int_{-\infty}^{0} \md k\,\hat c_{\sigma,k} \, \me^{\mi (k+k_F) x}.
\end{equation}
The low-energy excitations have momenta that are close to the Fermi
momentum,  $k \simeq \pm k_F$. Thus, we can formally extend the
integration limits to infinity and consider that we have two independent field
operators that each fulfill the standard anti-commutation relations.

By expressing the BCS Hamiltonian in terms of these field operators, two types of terms appear. On the one hand, there are resonant terms that couple left and right movers, and which oscillate slowly in real-space. On the other hand, there are off-resonant terms that couple left movers to left movers and right movers to right movers. These terms have phases $\me^{\pm 2 \mi k_F x}$ that oscillate fast compared to the relevant excitation energies. Their contribution is thus negligible and they can be discarded below. With singlet pairing, we then arrive at the Hamiltonian
\begin{equation}
\begin{split}
	\hat H_{\text{singlet}} =
	\Delta^*
		\int \md x\big[&
			\hat \psi_{L,\downarrow} (x)\hat \psi_{R,\uparrow}(x)
			\\+
			&\hat \psi_{R,\downarrow} (x)\hat  \psi_{L,\uparrow}(x)
		\big] 
		+
		\text{H.c.},
\end{split}
\end{equation}
with $\Delta \equiv \Delta_{\uparrow\downarrow}$.
As illustrated in Fig.~\ref{fig/dispersion}, this Hamiltonian couples $\hat \psi_{L,\downarrow}(x)$ to $\hat \psi_{R, \uparrow}(x)$ and $\hat \psi_{R, \downarrow}(x)$  to $\hat \psi_{L,\uparrow}(x)$ by the amplitude $\Delta$.

A similar derivation can be carried out for a triplet superconductor. 
For pairings of fermions with the same spin, the
Hamiltonian in momentum representation reads
\begin{equation}
	\hat H_{\text{triplet}} =
	\int
            \frac{\md k}{2 \pi}
	\left(
		\Delta^* \hat c_{-k} \hat c_{k}
		+
		\Delta \hat c^\dagger_{k} \hat c^\dagger_{-k}
	\right),
\end{equation}
where $\Delta\equiv\Delta_{\sigma \sigma}$ is now an odd function of the momentum~\cite{Sigrist_RMP}. 
In real-space, the Hamiltonian then becomes
\begin{equation}
	\hat H_{\text{triplet}} =
		\Delta^*
		\int \md x
			 \hat\psi_{L}(x) \hat\psi_{R}(x)
		+
		\text{H.c.}
\end{equation}

\subsection{Circuit description}

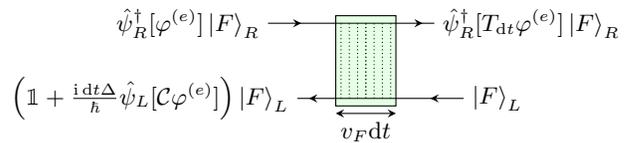
\begin{figure}
	\centering
	\begin{tikzpicture}
	\def\xmax{1.3}
	\def\ymax{0.5}
	\def\xsc{0.4}
	\begin{scope}[decoration={markings,
			mark=at position 0.2 with \arrow{stealth},
			mark=at position 0.93 with \arrow{stealth}
		}]
		\draw[postaction=decorate] (-\xmax, \ymax) -- (0.7*\xmax, \ymax)
			node[at start,left] {$\hat\psi_R^\dagger[\varphi^{(e)}]\ket{F}_R$}
			node[at end,right] {$\hat\psi_R^\dagger[T_{\md t}\varphi^{(e)}]\ket{F}_R$};
		\draw[postaction=decorate] (\xmax, -\ymax) -- (-0.7*\xmax, -\ymax)
			node[at end,left] {$\left(\mathbbm{1} + \frac{\mi \, \md t \Delta}{\hbar} \hat\psi_L[\mathcal{C}\varphi^{(e)}]\right) \ket{F}_L$}
			node[at start,right] {$\ket{F}_L$};
	\end{scope}

	\foreach \x in {0, 0.11, ..., \xsc} {
		\draw[densely dotted] (\x, -\ymax) -- (\x, \ymax);
		\if 0 \x\relax
		\else
			\draw[densely dotted] (-\x, -\ymax) -- (-\x, \ymax);
		\fi
	}
	\draw[fill=green,fill opacity=0.1] (-\xsc, 1.2*\ymax) rectangle (\xsc, -1.2*\ymax);

	\draw[stealth-stealth] (-\xsc, -1.4*\ymax) -- (\xsc, -1.4*\ymax)
		node[below,midway] {$v_F \md t$};
\end{tikzpicture}
	\caption{Andreev conversion on a thin superconductor.
		An incoming electron can either be transmitted through the superconductor or be Andreev reflected as a hole.
		The transmitted electron is delayed by the time $\md t = \md x/v_F$, where~$\md x$~is the width of the superconductor and $v_F$ is the Fermi velocity.
	}
\label{fig/andreev/elementary}
\end{figure}

We first consider the scattering of  electrons by a thin slice of a superconductor as in Fig.~\ref{fig/andreev/elementary}. Since
the Hamiltonian is local, the superconductor can be described by
stacking many of these small slices. Thus, we consider the scattering of
incoming electrons by a superconductor of thickness $\md x$. The
Hamiltonian being quadratic, the scattering process is linear. Furthermore, an incoming electron from the left will be scattered as an outgoing superposition of a left-moving hole and a right-moving electron. 

We aim to describe the state in the normal-metal region far from the
superconductor. Moreover, due to the linear dispersion, time and
position are equivalent, and we may express the state of an electron as
\begin{equation}
	\ket{\varphi^{(e)}} = v_F \int \varphi^{(e)}(v_F t) \ket{v_F t}  \md
	t,
\end{equation}
where $\varphi^{(e)}(x)=\langle x|\varphi^{(e)}\rangle$ is the wave
function of the electron, and we have used that $x=v_Ft$. This time representation will turn out to be very convenient. First of all, we may work with a fixed position far from any interfaces and instead vary time. Second, from a practical point of view, measurements are typically carried out at fixed positions and the data is acquired as a function of time or frequency. With this in mind, we will express the bras and kets in the time representation mainly.

If an incoming electron scatters off the superconductor, two processes
can occur. The first one is an Andreev conversion of the electron as a
hole, which happens with an amplitude that is proportional to the
thickness of the slice~$\md x$.
Since we are considering a thin slice of superconductor, much smaller
than the typical size of the wave packet, the response is expected to be constant over a wide range of frequencies. The state of the outgoing hole is then given by
$\mathcal{C}\ket{\varphi^{(e)}}$, where $\mathcal{C}$ is the anti-unitary
operator that describes the conversion of an electron into a hole. In
the time domain, this operator is given by complex conjugation
as~$(\mathcal{C} \varphi^{(e)})(t) = [\varphi^{(e)}(t)]^*$. It is,
however, more convenient to describe the conversion of an electron into
a hole in the frequency domain. If we define $\varphi^{(e)}(\omega)$ as
the Fourier transform of $\varphi^{(e)}(t)$, we see that
$(\mathcal{C}\varphi^{(e)})(\omega) = [\varphi^{(e)}(-\omega)]^*$. Thus,
the operator $\mathcal{C}$ transforms positive frequencies into negative
ones and vice versa without changing the wave packet, which corresponds
to the wideband assumption above.

The second process that can occur is normal transmission through the
superconductor, in which the outgoing electron is delayed by the time $\md
x/v_F$. We can express both of these processes using the second quantization. Hence, we
introduce the operators
\begin{equation}
	\hat\psi^\dagger_{L/R}[\varphi]
	=
	v_F \int \varphi(t) \hat\psi^\dagger_{L/R}(t) \md t
\end{equation}
that create left or right-moving electrons in the wave
function~$\varphi$. We can represent the scattering process as a linear transformation between creation and annihilation operators described by the Bogoliubov transformation
\begin{equation}
	\hat\psi^\dagger_R[\varphi^{(e)}]
	\to
	\hat\psi^\dagger_R[T_{\md x/v_F} \varphi^{(e)}]
	+ \mi \frac{\md x \Delta}{\hbar v_F} \hat\psi_L[\mathcal{C} \varphi^{(e)}],
	\label{eq/bogo/dx}
\end{equation}
where $T_{\mathrm{d} t}$ is an operator that translates time by the duration $\mathrm{d} t$.

Finally, one has to describe the ground state for the system composed by
the slice of superconductor surrounded by the two normal metals. 
This case is actually pretty simple since if the three subsystems are at zero temperature and at the same chemical potential they are at equilibrium. 
As such, an easy ground state consists of two incoming Fermi sea at the chemical potential of the superconductor.
The transformation in \cref{eq/bogo/dx} should then be applied on this state.
Hence, the processes in Fig.~\ref{fig/andreev/elementary} can occur if a single electron arrives on the left side of the superconductor, so that either an electron is transmitted through the superconductor or a hole is Andreev reflected.

\subsection{NS junction}
\begin{figure*}
	\centering
	    \begin{tikzpicture}
	\def\metal{2.5}
	\def\supra{2.5}
	\def\sep{0.2}
	\def\metalwidth{0.3}
	\def\letterheight{1.7}
	\pgfmathsetmacro{\axisheight}{-2.4*\metalwidth}
	\pgfmathsetmacro{\endx}{\metal+\supra+2*\sep}
	\begin{scope}[shift={(0,-\axisheight)}]
		\node at (-0.05*\endx, \letterheight+\axisheight) {(a)};
		\draw[->] (-0.05*\endx,\axisheight) -- +(1.1*\endx, 0)
			node[below] {$x$};

		\draw (0, \metalwidth) -- ++(\metal, 0)
			arc[start angle=90,end angle=-90,y radius=\metalwidth, x
			radius={0.3*\metalwidth}]
			-- ++(-\metal, 0);
		\draw (\metal, \metalwidth) arc[start angle=90,end angle=270,
			y radius=\metalwidth, x radius={0.3*\metalwidth}];

		\draw (\metal, 0) -- ++(2*\sep,0);

		\draw[pattern=north east lines] (\endx, \metalwidth) -- ++(-\supra, 0)
			-- ++(0, -2*\metalwidth) -- ++(\supra, 0);

		\draw[densely dashed] (\metal+\sep, 1.3*\metalwidth) --
			(\metal+\sep,\axisheight-0.3);

		\node[below left,font=\small] at (\metal+\sep, \axisheight) {Metal};
		\node[below right,font=\small] at (\metal+\sep, \axisheight) {Superconductor};

		\draw[white,thick,fill=blue] (0.65*\metal, \metalwidth+\sep)

			circle (0.07);
		\draw[-stealth] (0.65*\metal+0.15, \metalwidth+\sep) --
			++(0.5, 0);

		\draw[blue,thick,fill=white]
			(0.65*\metal+0.15+0.5, -\metalwidth-\sep) circle (0.07);
		\draw[-stealth]
			(0.65*\metal+0.5, -\metalwidth-\sep) -- ++(-0.5, 0);

		\begin{scope}[shift={(\endx-0.65*\supra-0.5-0.15, -\metalwidth-\sep)}]
			\fill[fill=blue] (0, 0.06) circle (0.05);
			\fill[fill=blue] (0, -0.06) circle (0.05);
			\draw (0, 0) ellipse[x radius=0.08, y radius=0.16];
			\draw[-stealth] (0.15, 0) -- ++(0.5, 0);
		\end{scope}
	\end{scope}

	\def\cellwidth{1.1}
	\def\cellheight{1}
	\def\numcells{2}
	\def\metal{1.5}
	\pgfmathsetmacro{\endxx}{\metal+(\numcells+0.4)*\cellwidth}
	\begin{scope}[shift={(\endx+3, 0)}]
		\node at (-0.05*\endxx, \letterheight) {(b)};
		\draw[->] (-0.05*\endxx,0) -- +(1.1*\endxx, 0)
			node[below] {$x$};

		\begin{scope}[shift={(0,-\axisheight-\metalwidth)}]
			\draw (0,0) -- (\endxx, 0);
			\draw (0,\cellheight) -- (\endxx,\cellheight);

			\foreach \i in {0,...,\numcells} {
				\fill (\metal+\i*\cellwidth, 0) circle (0.07);
				\fill (\metal+\i*\cellwidth, \cellheight) circle (0.07);
				\draw (\metal+\i*\cellwidth, 0) -- ++(0, \cellheight)
					node[midway,right=-2pt,font=\footnotesize] {$S_{\md x}$};
				\ifnum \i < \numcells
					\node[above=-2pt,font=\footnotesize]
						at ({\metal+(\i+0.5)*\cellwidth}, \cellheight)
						{$T_{\md t}$};
					\node[below=-1pt,font=\footnotesize]
						at ({\metal+(\i+0.5)*\cellwidth}, 0)
						{$T_{\md t}$};
				\fi
			}
			
			\draw[white,thick,fill=blue] (\metal-1, \cellheight+0.15)

				circle (0.07);
			\draw[-stealth] (\metal-1+0.1, \cellheight+0.15) --
				++(0.5, 0);

			\draw[blue,thick,fill=white]
				(\metal-0.4, -0.15) circle (0.07);
			\draw[-stealth]
				(\metal-0.5, -0.15) -- ++(-0.5, 0);
		\end{scope}
		\begin{scope}[shift={(\metal-0.2*\cellwidth,0)}]
			\draw[densely dashed] (0, -0.4) --
				++(0, 1.2+\cellheight);
			\node[below left,font=\small] at (0,0) {Metal};
			\node[below right,font=\small] at (0,0) {Superconductor};
		\end{scope}
	\end{scope}
\end{tikzpicture}
	\caption{Andreev conversion on a superconductor.
		(a) An incoming electron is reflected as a hole by the superconductor, where a Cooper pair is formed.
		(b) Inside the
		superconductor electrons can propagate ballistically, as described by the operator
		$T_{\md t}$, or be Andreev
		reflected, as described by the operator~$S_{\md x}$. Here, a small time-step is denoted by $\md t$, and $\md x = v_F \md t$.
	}
	\label{fig/andreev/model}
\end{figure*}
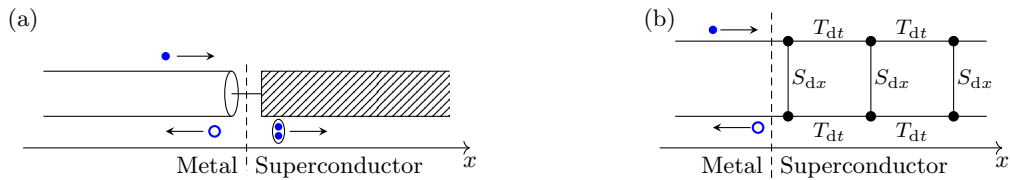

We now focus on the scattering off an interface between a normal-metal
and a superconductor. The central idea is to decompose the
semi-infinite superconductor into thin slices as shown in
Fig.~\ref{fig/andreev/model}. This approach is valid if we assume that the
coupling $\Delta$ does not depend on the wave vector~$k$ nor on the
position within the superconductor.

An incoming electron can be Andreev converted as a hole or it may
propagate inside the superconductor. The latter case corresponds to the
transmission of an electron from the metal into the superconductor, which can occur if the energy of the electron is above the gap.
For this reason, the outgoing state is not a pure state, but rather a
statistical mixture of a filled Fermi sea and a Fermi sea with a hole.
The state of the outgoing hole reads
\begin{equation}
	\sqrt{p} \ket{\varphi^{(h)}_{\text{out}}}
	=
	\mathcal{C} \mathcal{L} [\ket{\varphi^{(e)}_{\text{in}}}],
\end{equation}
where $p$ is the probability to have a conversion into a hole and
$\mathcal{L}$ is a linear operator that describes the response of the
superconductor, which we wish to determine.

To describe the NS interface we use the discrete element model depicted
in Fig.~\ref{fig/andreev/model}(b). Using the fact that the superconductor is semi-infinite and that the wave function should be smooth we find the following equation for the operator $\mathcal{L}$ (for the full derivation, see App.~\ref{app/deriveq/respNS})
\begin{equation}
	\frac{\Delta}{\hbar}\ket{\varphi^{(e)}_{\text{in}}}
	+ 2 \omega
	\mathcal{L}[\ket{\varphi^{(e)}_{\text{in}}}]
	+
	\frac{\Delta^*}{\hbar}
	\mathcal{L}^2[\ket{\varphi^{(e)}_{\text{in}}}]
	= 0.
 \label{eq:NS_resp}
\end{equation}
Since $\mathcal{L}$ is linear and invariant under time translations, we
can express its action in the frequency domain as
\begin{equation}
\mathcal{L}[\ket{\psi}](\omega) = \kappa(\omega) \psi(\omega)
\end{equation}
and the equation for the response function $\kappa(\omega)$ becomes
\begin{equation}
	\kappa^2(\omega) +
	2 
		\frac{\hbar \omega}{\Delta^*}
	\kappa(\omega)
	+ \frac{\Delta}{\Delta^*}
	=
	0.
\end{equation}
Moreover, writing the pairing potential as $\Delta = \me^{\mi
\varphi} |\Delta|$ and imposing that the response function is bounded
and causal we find the physical solution 
\begin{equation}
\kappa(\omega) = \me^{\mi \varphi} \Gamma(\omega), 
\label{eq:kappa_resp}
\end{equation}
where we have defined (with $x=\hbar\omega/|\Delta|$)
\begin{equation}
	\Gamma(x)
	=
	\begin{cases}
		x
		+ \sqrt{
			x^2
			-
			1
		},
		& \text{if $x<-1$}
		\\
		x
		- \mi \sqrt{
			1
			-
			x^2
		},
		& \text{if $-1 \leq x \leq  1$}
		\\
		x
		- \sqrt{
			x^2
			-
			1
		},
		& \text{if $x > 1$}
		\\
	\end{cases},
 \label{eq:gamma_resp}
\end{equation}
or, equivalently in the complex plane,
\begin{equation}
	\Gamma(z)
	=
	z
	- \sqrt{
		z
		+
		1
		+ \mi 0^+
	}
	\sqrt{
		z
		-
		1
		+ \mi 0^+
	}.
	\label{eq/NS/response}
\end{equation}
This particular solution has a single branch cut in the lower complex  half-plane for $z \in [-1 - \mi 0^+, 1 - \mi 0^+]$, which ensures that the response is causal. Indeed, we can return to the time-domain by an inverse Fourier transform,
\begin{equation}
	\Gamma(t)
	=
	\frac{1}{2\pi}
	\int_{-\infty}^\infty
	\Gamma(x)
	\me^{-\mi x t |\Delta|/\hbar }
	\md x.
\end{equation}
For $t < 0$, we close the contour in the upper half plane and the integral vanishes since no branch cuts or poles are enclosed. For $t > 0$, we instead evaluate the integral along the branch cut in the lower half plane and find
\begin{equation}
	\Gamma(t) = -\mi \Theta(t) \frac{J_1(|\Delta| t/\hbar)}{t},
\end{equation}
where $J_n(t)$ are the Bessel functions of the first kind and $\Theta(t)$ is a step function which ensures a causal response.

\subsection{NIS junction}
To describe the NIS junction we add an insulating layer that acts as a
tunnel barrier between the normal-metal and the superconductor. The
barrier can be described by its transmission and reflection amplitudes,
$t(\omega)$ and $r(\omega)$, which to a good approximation can be taken
to be frequency-independent. The setup resembles a Fabry-Pérot
interferometer with both the insulator and the superconductor
acting as reflecting mirrors~\cite{Kotilahti:2021}. We can then sum up the amplitudes for all
possible processes, and for a singlet superconductor we find
\begin{equation}
	r_{ee}(\omega)
	=
	r \frac{1 - \Gamma^2(\omega)}{1 - |r|^2\Gamma^2(\omega)},
\label{eq:r_ee_NIS_S}
\end{equation}
and
\begin{equation}
	r_{eh}(\omega)
	=
	{t^*}^2 \me^{\mi \varphi}
	\frac{\Gamma(\omega)}{1-|r|^2 \Gamma^2(\omega)},
\label{eq:r_eh_NIS_S}
\end{equation}
where we have used that $\kappa(\omega)\kappa^*(-\omega) =
-\Gamma^2(\omega)$ for a return trip inside the interferometer.

The case of a triplet superconductor is somewhat subtle because the
outgoing fields from the interferometer are the same. In that case, an
extra phase of $\pi$ is acquired during the return trip
within the interferometer and the reflection amplitudes then become
\begin{equation}
		r_{ee}(\omega) = r
			\frac{1 + \Gamma^2(\omega)}{1 + |r|^2\Gamma^2(\omega)},
   \label{eq:r_ee_NIS_T}
\end{equation}
and
\begin{equation}
		r_{eh}(\omega) = -{t^*}^2
			\me^{\mi\varphi}\frac{\Gamma(\omega)}{1 + |r|^2\Gamma^2(\omega)},
   \label{eq:r_eh_NIS_T}
\end{equation}
which have different signs compared to the singlet case.

\section{Wigner function}
\label{sec/coherence}

\subsection{First-order coherence}

To describe the electronic excitations, we consider the first-order coherence function defined as~\cite{bocquillon2014electron,Haack_2013,Moskalets_2015,Haack_2016,Moskalets_2016,moskalets2020composite,Kotilahti:2021}
\begin{equation}
	\mathcal{G}^{(e)}_{x,x'}(t, t')
	=
	\braket{\hat\psi^\dagger(x', t') \hat\psi(x, t)},
	\label{eq/coh/elec/def}
\end{equation}
where the position $x$ may also function as a channel and spin index. Typically, we consider a fixed position and omit this index. We also define the hole coherence as
\begin{equation}
	\mathcal{G}^{(h)}_{x,x'}(t, t')
	=
	\braket{\hat\psi(x',t') \hat\psi^\dagger(x, t)}.
	\label{eq/coh/hole/def}
\end{equation}
Both of the first-order coherences obey the Hermitian symmetry
$\mathcal{G}^{(e/h)}_{x,x'}(t, t') = [\mathcal{G}^{(e/h)}_{x', x}(t', t)]^*$. Also, because of the fermionic anticommutation relations, we have
\begin{equation}
	\mathcal{G}^{(e)}_{x, x'}(t, t')
	+
	\mathcal{G}^{(h)}_{x', x}(t', t)
	=
	\delta\left[
		(x - v_F t) - (x' - v_F t')
	\right]\, .
\end{equation}
In addition, one can define the
excess coherence, 
\begin{equation}
\Delta\mathcal{G}^{(e/h)}= \mathcal{G}^{(e/h)} -
\mathcal{G}^{(e/h)}_{\mu,T},
\end{equation}
where the equilibrium contribution $\mathcal{G}^{(e/h)}_{\mu,T}$ from the Fermi sea at chemical potential~$\mu$
and temperature~$T$ has been subtracted. We may then express the electric current as
\begin{equation}
	i(t)
	=
	- e v_F \Delta\mathcal{G}^{(e)}(t, t),
\end{equation}
which vanishes at equilibrium where $\mathcal{G}^{(e/h)} =
\mathcal{G}^{(e/h)}_{\mu,T}$.

\subsection{Wigner representation}
\label{sec/wignerrep}
The first-order coherences in Eqs.~(\ref{eq/coh/elec/def},\ref{eq/coh/hole/def}) provide information about the electronic state in the time-domain. To visualize this information, we here adopt the time-frequency Wigner
representation defined as~\cite{ferraro2013wigner,Ferraro:2014,Roussel:2021}
\begin{equation}
	W(t, \omega)
	=
	v_F \int \mathcal{G}(t + \tau/2, t-\tau/2)
	\me^{\mi \omega \tau} \md \tau.
 \label{eq:Wig-func-def}
\end{equation}
While the Wigner function is real, it is in general not positive, nor
bounded from above by one. For electrons, values below zero or above one indicate non-classicality.
We also introduce the excess Wigner function,
\begin{equation}
\Delta W = W -
W_{\mu, T},
\end{equation}
where we again subtract the contribution from the Fermi sea. The current can then be written as
\begin{equation}
	i(t) = -e \int \Delta W^{(e)}(t, \omega) \, \frac{\md \omega}{2 \pi}, 
\end{equation}
while the excess occupation number at energy $\hbar\omega$ reads
\begin{equation}
	\Delta n(\omega) = \overline{\Delta W^{(e)}(t, \omega)}^t.
\end{equation}
Here, the overline denotes an average over a period
for periodic drives and an integral over time for non-periodic drives. The anticommutation relations imply that
\begin{equation}
	W^{(e)}(t, \omega) + W^{(h)}(t, -\omega) = 1, 
\end{equation}
and
 \begin{equation}
	\Delta W^{(e)}(t, \omega) + \Delta W^{(h)}(t, -\omega) = 0.
\end{equation}

Wigner representations are also useful to visualize the scattering
process in itself. First of all, both linear and antilinear
transformations acting on the wavefunctions can be represented by linear
transformations on the Wigner functions. Furthermore, the
transformations that are invariant by time translations have a
particularly simple form and can be associated with a Wigner function.

First of all, any antilinear transformation can be
represented as the combination of a linear transformation and the
electron-hole transformation given by the~$\mathcal{C}$ operator
introduced earlier. The latter transforms electrons into holes and vice versa, corresponding to the
transformation that maps $\Delta W^{(e)}$ to $\Delta W^{(h)}$ or, equivalently,
\begin{equation}
	\Delta W(t, \omega) \mapsto - \Delta W(t, -\omega) .
\end{equation}
In addition, any time-invariant linear scattering process that transforms the
operator $\hat\psi(t) \to \int \kappa(t-t') \hat\psi(t')  \md t'$, where $\kappa$ is the response function, can be
associated to the Wigner function of the response, defined as
\begin{equation}
W_\kappa(t, \omega) = \int
\kappa(t+\tau/2) \kappa^*(t-\tau/2) \me^{\mi \omega \tau}  \md \tau.
\end{equation}
The  Wigner function of the outgoing state is then given by the incoming one convolved with the response as~\cite{flandrin1998time}
\begin{equation}
	W^{\text{(out)}}(t, \omega)
	=
	\int W_\kappa(t-t', \omega) W^{\text{(in)}}(t', \omega)
	\md t .
\end{equation}
As such, the Wigner function $W_\kappa$ provides a convenient
visualization of the response function. Its marginals contain
information about the response in the time and in the frequency domain.
The marginal in frequency yields the probability of a wave packet with a
given  frequency to be transmitted. Similarly, the marginal in time
yields the modulus squared of the response, $|\kappa(t)|^2$. It is also
possible to extract the time delay of a wave packet at a given
frequency. If a wave packet is peaked around the frequency
$\omega$, with a width that is small compared to the variations of the
Wigner function of the response, it will experience the  time delay~\cite{flandrin1998time}
\begin{equation}
	\tau(\omega)
	=
	\int t  W_\kappa(t, \omega)  \md t
	=
	\mathrm{Im}[\kappa'(\omega)/\kappa(\omega)].
\label{eq:delaytime}
\end{equation}

\section{Response function}
\label{sec/respfunc}

\begin{figure}
\includegraphics[width=0.95\columnwidth]{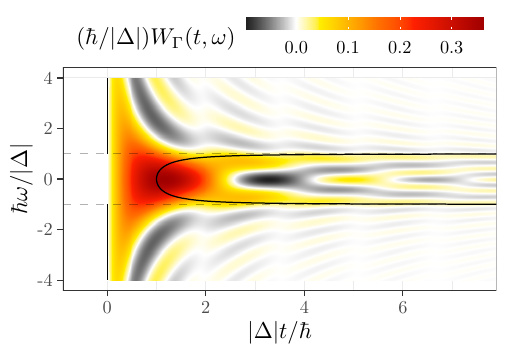}
	\caption{
		Wigner representation of the response function of an
		NS junction.
		The gap is marked by the dashed line while the black line represents the delay time~$\tau_\text{NS}(\omega)$.
	}
	\label{fig/respwig}
\end{figure}

\subsection{NS junction}

We now investigate the response functions of the NS and NIS junctions. In
general, we have four different responses, two for normal
electron-electron and hole-hole processes, and two for electron-hole
and hole-electron conversion processes. However, due to the
electron-hole symmetry, we can focus on just two of them, for example,
the normal electron-electron processes and the Andreev electron-hole conversion
processes. We start with the NS junction, for which there are
no normal reflections, and it suffices to consider the conversion
processes only. 

The response of the NS junction is given by Eqs.~(\ref{eq:kappa_resp},\ref{eq:gamma_resp}). Thus, an incoming electron with the
wave function $\varphi^{(e)}(\omega)$ will be converted into a hole with the
wave function $\varphi^{(h)}(\omega)$ according to the relation
\begin{equation}
	\sqrt{p} [\varphi^{(h)}(-\omega)]^*
	=
	\Gamma(\omega) \varphi^{(e)}(\omega).
\label{eq:freq_resp_NS}
\end{equation}
Here, the probability of conversion can be expressed as
\begin{equation}
	p =
	\int_0^\infty
	|\Gamma(\omega) \varphi^{(e)}(\omega)|^2
	\md \omega,
\end{equation}
which ensures that $\varphi^{(h)}$ is properly normalized. We can also revert Eq.~(\ref{eq:freq_resp_NS}) to the time-domain  as
\begin{equation}
	\sqrt{p}\varphi^{(h)}(t)^*
	=
	\int \Gamma(t-t') \varphi^{(e)}(t') \md t',
\end{equation}
showing how the incoming wave packet is modified.

To analyze the conversion probability at frequency~$\omega$, we notice
that $\Gamma(\omega)$ in Eq.~(\ref{eq:gamma_resp}) takes the form
$\Gamma(\omega)=\me^{-i\arccos{(\hbar\omega/|\Delta|)}}$ for
$|\hbar\omega|\leq|\Delta|$.  Thus, for a wave packet with energies
inside the gap the conversion probability is one. However, at higher
energies, the probability of conversion decays. Indeed, there is a
chance that an electron containing energies above the gap escapes into
the superconductor.
Conversely, a simple case is when the energies of the incoming wave
packet are well inside the gap, such that
$|\hbar\omega|\ll|\Delta|$. Since we have $\Gamma(\omega) \simeq
- \mi \me^{\mi \hbar \omega/|\Delta|}$ the outgoing wave
packet is not deformed but only delayed by the time $\tau_{\text{NS}}
=\hbar/|\Delta|$. More generally, we find the delay time at
frequency~$\omega$
\begin{equation}
	\tau_{\text{NS}}(\omega)
	=
	\frac{\hbar}{\sqrt{|\Delta|^2 - (\hbar \omega)^2}}
 \label{eq:NS_delay}
\end{equation}
for energies inside the gap according to Eq.~(\ref{eq:delaytime}). Here
we see that the delay time diverges as we approach the superconducting
gap. By contrast, above the gap the delay time is zero and the
scattering process is instantaneous. 

These observations can be visualized by the Wigner function in
Fig.~\ref{fig/respwig}. Clearly, the response is causal, as one would
expect since $W(t, \omega) = 0$ for $t < 0$. We also indicate the delay
time given by \cref{eq:NS_delay} (black line), which is related to the Wigner function
according to Eq.~(\ref{eq:delaytime}).
In the Wigner function of the response, we see a difference in
the oscillations above and below the gap. In
particular, above the gap the positive and
negative parts tend to balance out. Indeed, because the Wigner function is
causal, and the time delay above the gap is zero, we have
\begin{equation}
	\int_0^\infty W(t, \omega) t \, \md t = 0 \, ,
\end{equation}
which causes the positive and negative parts to cancel out. By contrast,
inside the gap the time delay is finite and it even diverges close to
the gap, which explains the different oscillatory behaviors above and below the gap.

The interpretation of
$\tau_{\text{NS}}(\omega)$
as a delay time is only valid for wave functions that are centered around
$\omega$ with a small bandwidth compared to the frequencies over which
the time delay changes. Conversely, the duration of the
wave packet must be large compared to the oscillations in the Wigner
function at this frequency. 
Hence, for frequencies close to the gap edge in \cref{fig/respwig}, where the delay time diverges, the wave functions must be very wide in order not to change shape due to the scattering.

\begin{figure}
\includegraphics{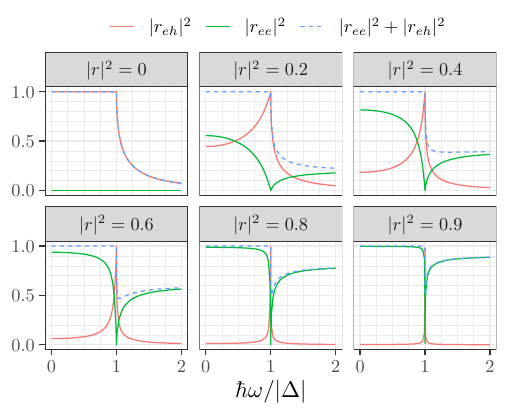}
	\caption{%
		Probabilities for Andreev conversions and normal-reflections on an NIS junction with a singlet
		superconductor.
	}
	\label{fig/r/singlet/modulus}
\end{figure}

\subsection{NIS junction -- singlet superconductor}

\begin{figure*}
	\includegraphics{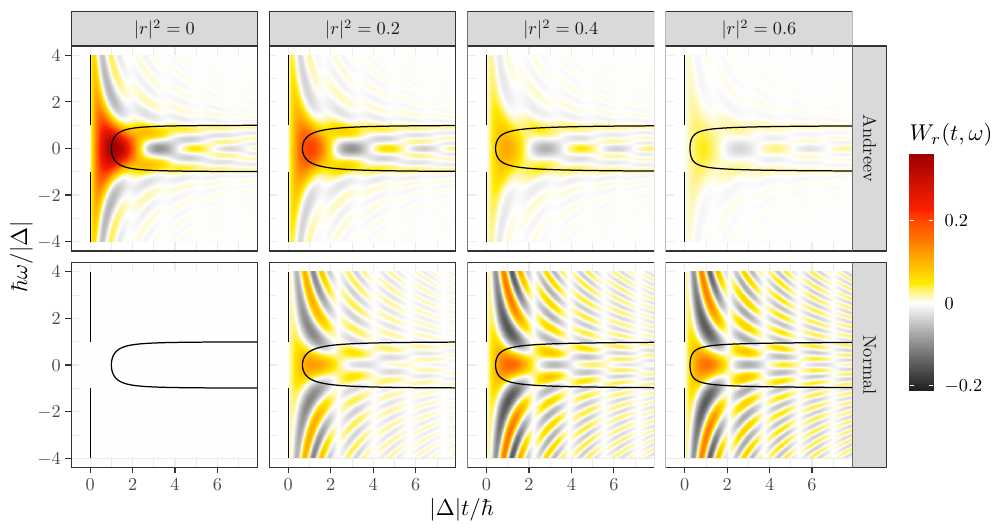}
	\caption{
		Wigner representation of the response function for an NIS
		junction with a singlet superconductor. The top row shows the Andreev conversions and the bottom row the normal-reflections. The black line indicates the delay time~$\tau_S(\omega)$.
	}
	\label{fig/respwig/sing}
\end{figure*}

We now consider the NIS junction, focusing first on singlet
superconductors. For the NIS junction, incoming electrons can be
converted into holes with the amplitude $r_{eh}(\omega)$ according to
Eq.~(\ref{eq:r_eh_NIS_S}). However, due to the insulating barrier, they
may also be reflected as electrons with amplitude $r_{ee}(\omega)$
according to Eq.~(\ref{eq:r_ee_NIS_S}). Thus, we have to consider both
processes in our analysis.

We first analyze the amplitudes at a given frequency.
For energies within the gap, electrons are never transmitted into the
superconductor and we have
$|r_{ee}(\omega)|^2 + |r_{eh}(\omega)|^2 = 1$. Above the
gap, electrons can be transmitted into the superconductor and the sum
is smaller than one. At very high energies, the amplitude
for electron-hole conversion vanishes and electrons are either
reflected back into the edge channel with probability $|r|^2$ or they are transmitted into the superconductor with
probability $1-|r|^2$. As such, we can
interpret $|r|^2$ as the probability of normal reflections at very high energies.

Figure~\ref{fig/r/singlet/modulus} shows the probabilities of normal reflections and
Andreev conversions for different reflection probabilities of the insulating barrier. The
conversion probability has a
local minimum at zero energy. Furthermore, there is a resonance peak  at the
gap edge where the probability for electron-hole conversion is one.
Moreover, when the reflection probability of the insulating barrier
approaches one a discontinuity develops around the gap for the sum
$|r_{ee}|^2+ |r_{eh}|^2$. This surprising feature implies that, if a
wavepacket is emitted slightly above the gap, its probability of
escaping into the superconductor can increase as $|r|^2$ increases.
At low energies, we can expand the amplitudes to first order in
$\hbar\omega/|\Delta|$ as
\begin{equation}
	r_{ee}(\omega)
	\simeq
	\frac{2 r}{1 + |r|^2}
	\left(
		1 + \mi \frac{\hbar \omega}{|\Delta|}
		\frac{1 - |r|^2}{1 + |r|^2}
	\right),
\end{equation}
 and
 \begin{equation}
	r_{eh}(\omega)
	\simeq
	-\mi \frac{{t^*}^2 \me^{\mi \varphi}}{1 + |r|^2}
	\left(
		1 + \mi \frac{\hbar \omega}{|\Delta|} \frac{1 - |r|^2}{1 + |r|^2}
	\right).
\end{equation}
Thus, at zero energy the normal reflection probability is $4 |r|^2/(1 +
|r|^2)^2$, while the Andreev conversion probability reads $(1 - |r|^2)^2/(1 +
|r|^2)^2$. The interference due to the barrier leads to an enhancement
of the normal reflections while it reduces the Andreev conversions.

For both processes, the time delay at low energies reads
\begin{equation}
\tau_S(\omega) \simeq \frac{\hbar}{|\Delta|} \frac{1 -
|r|^2}{1 + |r|^2}.
\end{equation}
Surprisingly, the time delay decreases with increasing reflection
probability and it vanishes as the reflection probability approaches
one. It must be noted that this happens only when the conversion
processes have some ``losses'', either in the form of trivial
reflection or through leakage in the superconductor, when the electron
is emitted above the gap. The fact that the conversion process becomes
faster than the typical time, $\hbar/\Delta$, or even instantaneous when
the electron is emitted above the gap, may seem to violate causality.
However, in lossy system, this is a relatively common occurence, and it
can be seen as a destructive interference
process, that does not violate causality~\cite{landauer1994barrier,winful2006tunneling,yao2012frequency}.

At low energies, we may also express the outgoing current in terms of the incoming one as
\begin{equation}
	i_{\text{out}}(t)
	\simeq
	\left(\frac{8 |r|^2}{(1+|r|^2)^2} - 1\right)
	i_{\text{in}}(t).
\end{equation}
From this expression we see that the current vanishes for $|r|^2 = 3 - \sqrt{8}
\simeq 0.17$, implying that the number of reflected electrons and holes exactly cancel out in that case. Interestingly, this cancellation occurs well below $|r|^2 = 0.5$ because of the reduction of the electron-hole conversion processes due to the barrier.

\Cref{fig/respwig/sing} shows the Wigner representation of the conversion processes and the reflection processes for  different probabilities of reflections on the barrier. 
Increasing the reflection probability, the conversion processes are reduced and the reflection of electrons on the interface becomes increasingly dominant. 
In addition, we notice a resonance in the conversion process at energies close to the gap. In the figure, we also show the delay time for energies inside gap, which can be written as
\begin{equation}
	\tau_S(\omega)
	=
	\tau_{\text{NS}}(\omega) \left(
		1
		-
		2 |r|^2 \frac{
			(1 + |r|^2)
			-
			2 ( \hbar \omega/|\Delta| )^2
		}{
			(1 + |r|^2)^2
			- 4 |r|^2 
			( \hbar \omega/|\Delta| )^2
		}
	\right),
\end{equation}
where $\tau_{\text{NS}}(\omega)$ is the delay time for the NS junction. From this expression we see that the delay is decreased compared to the NS junction for small energies, $|\hbar\omega/\Delta| < \sqrt{(1 + |r|^2)/2}$, while it is increased for larger energies.

\subsection{NIS junction -- triplet superconductor}

\begin{figure}
\includegraphics{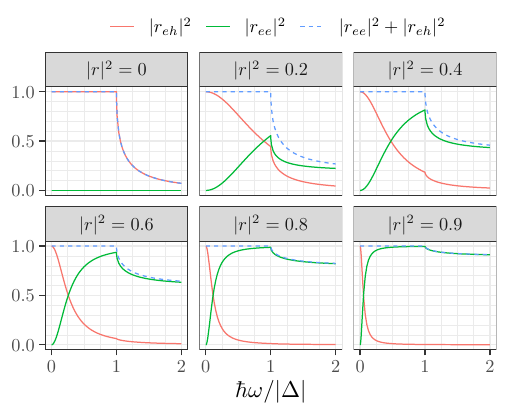}
	\caption{%
		Probabilities for Andreev conversions and normal-reflections on an NIS junction with a triplet
		superconductor.
	}
	\label{fig/r/triplet/modulus}
\end{figure}

\begin{figure*}
	\includegraphics{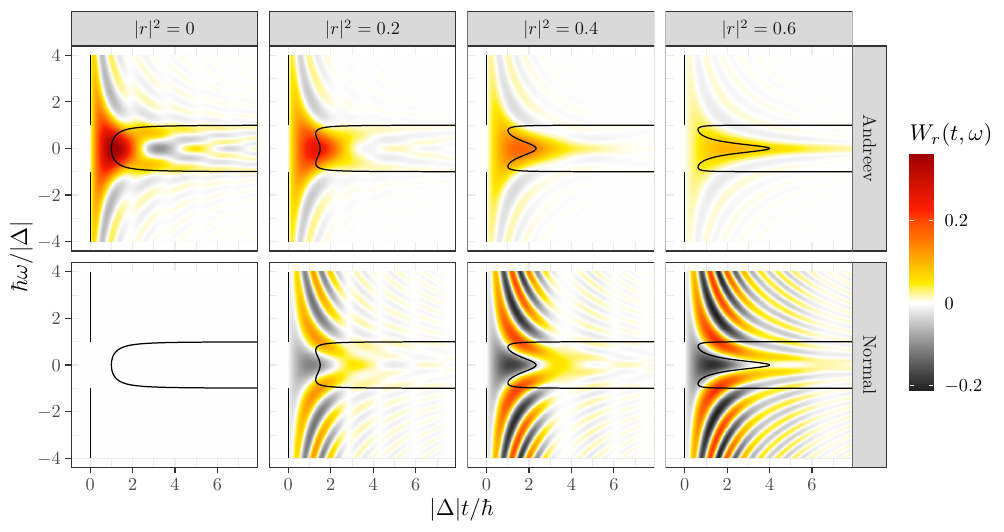}
	\caption{
		Wigner representation of the response function for an NIS
		junction with a triplet superconductor. The top row shows the Andreev conversions, while the bottom row shows the normal-reflections. The black line indicates the delay time~$\tau_T(\omega)$.
	}
	\label{fig/respwig/trip}
\end{figure*}

We can carry out a similar analysis for the NIS junction with a triplet
superconductor. Similarly to the singlet superconductor, at energies well above
the superconducting gap incoming electrons are either normal 
reflected with probability
$|r|^2$ or they are transmitted into the superconductor with probability $1-|r|^2$; moreover, there is no Andreev conversion.
At lower energies, the triplet and the singlet cases are
different because of the different signs between
Eqs.~(\ref{eq:r_ee_NIS_S},\ref{eq:r_eh_NIS_S}) for a singlet
superconductor and Eqs.~(\ref{eq:r_ee_NIS_T},\ref{eq:r_eh_NIS_T}) for a
triplet superconductor. 

In \cref{fig/r/triplet/modulus} we show the normal reflection
and Andreev conversion probabilities for the triplet superconductor, which can
be compared with the results for the singlet superconductor in
\cref{fig/r/singlet/modulus}. First, unlike the singlet case, the
sum $|r_{ee}|^2+|r_{eh}|^2$ is smoother as $|r|^2$ approaches one.
Hence, transmissions into the superconductor are always reduced as $|r|^2$ increases.
We also notice that there is
always perfect conversion at zero energy. The conversion probability
then goes down with increasing energy and with increasing reflection
amplitude of the barrier. In particular, with a large reflection
amplitude the conversion process develops a narrow peak around zero
energy. At low energies, we can expand the amplitudes to first order in
$\hbar\omega/|\Delta|$ as
\begin{equation}
	r_{ee}(\omega)
	\simeq
	-2 \mi \frac{r}{1 - |r|^2} \frac{\hbar \omega}{|\Delta|},
\end{equation}
and
\begin{equation}
	r_{eh}(\omega)
\simeq
	\mi \frac{{t^*}^2}{|t|^2} \me^{\mi \varphi} \left(
		1 + \mi \frac{\hbar \omega}{|\Delta|}
		\frac{1 + |r|^2}{1 - |r|^2}
	\right).
\end{equation}
For normal reflections we then have $|r_{ee}(\omega)|^2
\simeq 4 |r|^2 (\hbar\omega/|\Delta|)^2$, which vanishes quadratically
with the energy. For the conversion processes we find at low energies  
\begin{equation}
\tau_T(\omega) \simeq \frac{\hbar}{|\Delta|}\frac{1 + |r|^2}{1 -
|r|^2},
\end{equation}
which is larger than $\hbar/|\Delta|$ and diverges as the reflection probability approaches one. For the expansions above to be valid the energies must be such that $|\hbar \omega | \ll |\Delta|, (1-|r|^2) |\Delta/r|$. Thus, the
frequency bandwidth for which these expansions hold decreases as the 
reflection probability goes to one. In turn, the delay time stays small
compared to the typical width of an incoming wave packet.

Figure~\ref{fig/respwig/trip} shows the Wigner representations of the
two processes for different reflection probabilities of the barrier. As
the reflection probability increases, the Andreev conversion processes
are reduced and the normal reflections become more pronounced. Still,
the behavior is different from the singlet case. For the triplet case,
we observe a resonance peak for the conversion processes at zero energy,
while in the singlet case the resonance occurs at the gap edge.
Moreover, the conversion processes for the singlet case are suppressed
at low energies.

Also, for the triplet case we can express the delay time for energies inside the superconducting gap as
\begin{equation}
		\tau_T(\omega)
	=
	\tau_{\text{NS}}(\omega) \left(
		1
		+
		2 |r|^2 \frac{
			(1 - |r|^2)
			-
			2 \left( \hbar \omega/|\Delta| \right)^2
		}{
			(1 - |r|^2)^2
			+
			4 |r|^2 
			\left( \hbar \omega/|\Delta| \right)^2
		}
	\right),
\end{equation}
where $\tau_{\text{NS}}(\omega)$ is the delay time for the NS junction. This expression is also different from the singlet case. At low
energies, $|\hbar\omega/\Delta| \leq \sqrt{(1 - |r|^2)/2}$, the time delay
is longer than for the NS junction; otherwise, it is shorter.

\section{Andreev conversion of a Leviton}
\label{sec/leviton}

\subsection{Levitons}

We can now consider the reflection and conversion of
specific charge excitations. In particular, we are interested in periodic voltage
pulses that excite an average charge  of
exactly one electron per period of the
drive~\cite{Levitov:1996,Ivanov:1997,safi1999dynamic,Keeling:2006,Dashti:2019,Moskalets_book}. If the
pulses are well-separated and lorentzian-shaped, they generate clean single-electron excitations known as levitons without exciting any additional electron-hole pairs in the Fermi
sea~\cite{dubois2013minimal,jullien2014quantum}. The injected
wave function reads
\begin{equation}
	\psi(t) =
	\sqrt{\frac{\tau_0}{v_F\pi}}
	\frac{1}{t - \mi \tau_0},
\end{equation}
whose modulus square is a lorentzian of width $\tau_0$. At zero temperature, we have $\Delta\mathcal{G}^{(e)}(t, t') = \psi^*(t') \psi(t)$ and the excess coherence of a leviton becomes
\begin{equation}
	\Delta\mathcal{G}^{(e)}(t, t')=\frac{\tau_0}{v_F\pi}\frac{1}{(t-i\tau_0)(t'+i\tau_0)}.
\end{equation}
Inserting this expression into Eq.~(\ref{eq:Wig-func-def}) and evaluating the integral using complex-contour integration we find
\begin{equation}
	\Delta W^{(e)}(t, \omega)
=8 \omega \tau_0\sinc(2\omega t)\me^{-2\omega\tau_0}\Theta(\omega) ,
\end{equation}
which is the Wigner representation of a leviton. Moreover, the current
is a lorentzian of width $\tau_0$,
\begin{equation}
	i(t)
	=
-\frac{e\tau_0}{\pi(t^2+\tau_0^2)}.
\end{equation}
We also find that the excess occupation number is
\begin{equation}
	\Delta n(\omega) = 4\pi\tau_0\me^{-2\omega\tau_0}\Theta(\omega),
\end{equation}
showing that these excitations have an exponentially decaying
distribution of energies. In Fig.~\ref{fig/wigner/lev/ns}, we show the
Wigner representation of a leviton. In the following, it will be important to compare the width of the wave function with the superconducting gap since their ratio will determine to what extent the wave packet is inside the gap, or if it also contains energies above the gap.

\subsection{NS junction}
\begin{figure*}
	\includegraphics{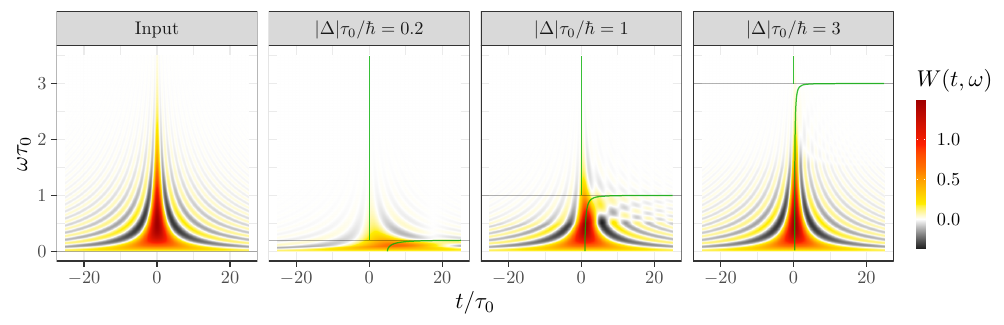}
	\caption{
		Andreev reflection of a leviton on an NS junction.
		The left panel  shows the electronic Wigner
		function~$W^{(e)}(t, \omega)$ of an incoming leviton of width~$\tau_0$.
		The other panels show the hole Wigner
		function~$W^{(h)}(t, \omega) = 1 - W^{(e)}(t, -\omega)$ of an Andreev converted leviton of different widths. The gray lines
		indicate the gap and the green lines show the delay
		time~$\tau_{\mathrm{NS}}(\omega)$.
	}
	\label{fig/wigner/lev/ns}
\end{figure*}

For the NS junction, the only possible process is the Andreev
conversion. In Fig.~\ref{fig/wigner/lev/ns}, we show the electronic Wigner
representation of an incoming leviton pulse  and the hole
Wigner representation of the outgoing holes for an NS junction. We consider different widths of the charge pulses compared to the gap of the superconductor. For the
outgoing holes, we use the hole representation $W^{(h)}(t, \omega) =
1-W^{(e)}(t, -\omega)$, which makes it easier to compare the Wigner functions for the incoming and the
outgoing particles. For the outgoing particles, we also
show the delay time~$\tau_{\text{NS}}(\omega)$ with a green line.

The converted holes change shape depending on the width of the incoming leviton. If the incoming leviton is well
within the gap, the converted hole is
an anti-leviton that is delayed by the time $\hbar/|\Delta|$. That situation corresponds to
the right-most panel in
\cref{fig/wigner/lev/ns}. We also see that 
the delay time $\tau_{\text{NS}}(\omega)$ is approximately equal to
$\hbar/|\Delta|$ over the full frequency range of the leviton. Furthermore, the conversion process happens with a
probability close to one. On the other hand, as the width of the leviton
is decreased, the conversion happens with a probability that is smaller
than one, since parts of the leviton may be transmitted into the
superconductor. In that case, the reflected particle is no longer an
anti-leviton as seen in the two central panels. If the frequencies of
the incoming leviton become comparable to the gap,  different frequency
components experience different delays, which changes the shape of the
reflected particles.
We also notice the consequence of the divergence of the time delay
around the gap. In the third panel of \cref{fig/wigner/lev/ns} the wave function
is strongly deformed around these frequencies. By contrast, at energies
well inside the gap the wave function is hardly changed, besides a time
delay, since the Wigner function of the leviton oscillates slowly. In the second panel of
\cref{fig/wigner/lev/ns}, the probability  for transmission into the
superconductor is large. In this case, the leviton duration is extremely
short compared to all the relevant scales of the problem. As such, the leviton becomes a wideband probe as the  Wigner function of the reflected particle resembles
the Wigner function of the response of the superconductor.

\subsection{NIS junction}

For the NIS junction, we have to consider both the Andreev conversions and the
normal reflections. For that reason, both
the time delay and the probability of reflections at a given frequency
play a role, even within the gap. Inside the gap, the number of excitations
(electrons and holes) is conserved and the conversion
and reflection processes are complementary: If the probability of one of the processes decreases, it increases for the other.

\begin{figure*}
	\includegraphics{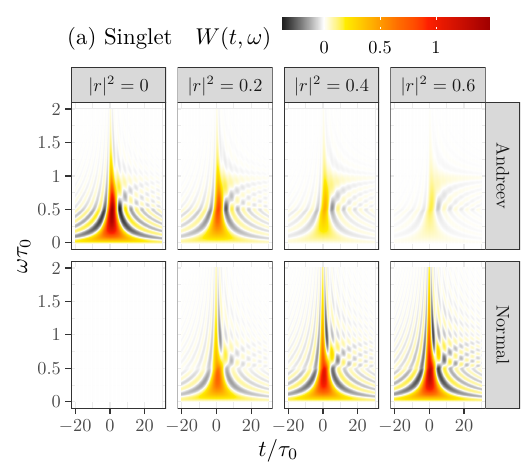}
	\hfill
	\includegraphics{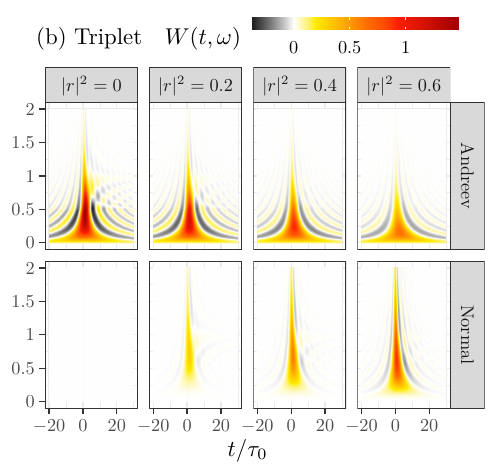}
	\caption{Wigner functions for  Andreev conversions and normal-reflections of a leviton. The Andreev conversions  (top row) are represented by the hole Wigner function~$W^{(h)}(t, \omega)$, while the normal reflections (bottom row) are represented by the
		electronic Wigner function~$W^{(e)}(t, \omega)$. The two panels
	correspond to singlet (a) and triplet (b) superconductors. In each
panel, we show results for different reflection probabilities of the
insulating barrier. The width of the incoming leviton is fixed to $
\tau_0 = \hbar/|\Delta|$.}
	\label{fig/lev/wig/nis/all}
\end{figure*}

In~\cref{fig/lev/wig/nis/all}, we show the results for a normal-reflected and an Andreev converted leviton, both for singlet and triplet superconductors. To simplify the comparison of the Andreev conversions and the normal-reflections, we show the Wigner function of the conversion processes as the Wigner representation of the hole coherence, $W^{(h)}(t, \omega) = 1 - W^{(e)}(t, -\omega)$. On the other hand, the Wigner function of the reflections is given by the electron coherence,~$\Delta W^{(e)}(t, \omega)$. In this way, we can directly compare the converted and the reflected 
wave packets.

We first observe that the normal processes are favored over the Andreev
processes as the reflection probability of the insulating barrier is
increased, both for singlet and triplet superconductors. Still,
comparing the  singlet and triplet superconductors for a given
reflection coefficient of the barrier, the singlet superconductor tends
to favor the normal processes over the Andreev processes more than the
triplet superconductor. We also see  an attenuation of the Wigner
functions at energies above the gap for the Andreev processes, which
corresponds to transmission of excitations into the superconductor. By
contrast, the normal processes do not exhibit this feature, which is
consistent with the fact that they are simply given by reflections on
the insulating barrier.

Altogether, the Wigner function visualizes the full
frequency dependence of the two processes and it allows for an accurate comparison
between singlet and triplet superconductors. For singlet superconductors, we see a long positive tail around
the gap for the Andreev processes, corresponding to the resonance of the
Andreev conversion at gap edge. As the barrier reflection increases, we observe
that the tail becomes better defined and it grows longer, corresponding to a
sharper energy filtering. Conversely, for the normal processes we see that
the Wigner function oscillates quickly between positive and negative
values around the gap edge, such that those frequencies are filtered out.
Indeed, in the triplet case there is no such resonance around the gap.
However, we observe that the low-frequency content of the Andreev
conversions remains important even at high values of the barrier reflection because
of the resonance at zero energy. Furthermore, we see that the center of the wave packet at low energies is  delayed as the barrier reflection increases. Conversely, the normal processes do not have any low-frequency components.

\section{Conclusions}
\label{sec/conclusion}

We have employed a Wigner function representation to describe the
scattering of single-electron charge pulses on the interface between a normal-metal and a
superconductor.
In particular, we have shown that Wigner functions provide an efficient
way of visualizing both the response of the superconducting interface
and the wave function of the outgoing particles thanks to the mixed time-and-frequency representation. From the Wigner functions several other quantities of interest can be extracted such as the time-dependent current, the probability for a charge at a particular frequency to be Andreev converted or normal-reflected at the interface, as well as the time delay associated with these processes.

We have considered junctions with and without an insulating barrier between the metal and the superconductor, which can have either singlet or triplet pairing. Without a  barrier, there are no normal reflections at the interface and incoming charges are either Andreev converted or transmitted into the superconductor above the gap. With the insulating barrier at the interface, more processes become possible as incoming charges may also be normal-reflected at the interface. The Wigner function representation allows us to visualize how an incoming wave packet is deformed by the scattering off the interface with the superconductor. Specifically, we have shown how the type of superconductor, being singlet or triplet, affects the response of the junction, for example by changing the probabilities of normal reflections at a given frequency and the associated time delay. These effects can be clearly seen in the Wigner function of Andreev converted and normal-reflected charge pulses. 

Our work shows that techniques from electron
quantum optics can be useful to explore the interface between superconductors and 
quantum Hall edge states. For example, by using time-dependent voltages one may obtain information beyond what is available from measurements with constant biases only~\cite{zhang2019perfectcar,kurilovich2022disorder}. Beyond exploring the physics of such interfaces, superconductivity may enable new applications in electron quantum optics such as the conversion of electrons into holes~\cite{aBurset_short}, which is a process without any counterpart in conventional quantum optics with photons.
Superconductors may also function as sources of spin entanglement by splitting the Cooper pairs in a superconductor into separated normal-state regions~\cite{Pandey:2021,Wang:2022,Bordoloi:2022,Bertin:2023a}. Furthermore, in terms of metrological
applications of electron quantum optics,
superconductivity may offer new possibilities, for instance, for the interferometric sensing of tiny magnetic fields.

\section{Acknowledgements}

We thank M.~V.~Moskalets for useful discussions and acknowledge support from InstituteQ, Research Council of Finland through Finnish Centre of Excellence in Quantum Technology (project No.~352925), 
the Spanish CM ``Talento Program'' project No.~2019-T1/IND-14088 and No.~2023-5A/IND-28927, the Agencia Estatal de Investigaci\'on project No.~PID2020-117992GA-I00 and No.~CNS2022-135950 and through the ``María de Maeztu'' Programme for Units of Excellence in R\&D (CEX2023-001316-M).

\appendix

\section{Derivation of response function}
\label{app/deriveq/respNS}
To derive the response function of the NS junction, we consider a linear operator $\mathcal{L}_x$ that transforms the incoming wave function at position $x$ into the outgoing one as
\begin{equation}
	\ket{\varphi^{(h)}_\text{out}(x)}
	=
	\mathcal{C} \mathcal{L}_x \ket{\varphi^{(e)}_\text{in}(x)},
 \label{eq:in-out_app}
\end{equation}
where $\mathcal{C}$ is an anti-unitary operator that describes the conversion of an electron into a hole.
In addition, the wave functions at $x$ and $x + \md x$ are related as
\begin{equation}
\begin{split}
	\mathcal{C}\ket{\varphi^{(h)}_\text{out}(x)}
	=&
	- S_{eh}^*\ket{\varphi^{(e)}_\text{in}(x)}\\
	&+
	S_{ee}^* T_{\md x/v_F} \mathcal{C}\ket{\varphi^{(h)}_\text{out}(x +\md x)},
\end{split}
\end{equation}
which follows from Fig.~\ref{fig/andreev/model}. The operator  $T_{\delta t}$ translates time by $\delta t$, while $S_{eh}$ describes Andreev reflections, and $S_{ee}$ accounts for transmissions into the superconductor.

Since the junction is situated at $x=0$, and the superconductor is considered to be semi-infinite, we have $\mathcal{L}_{x} =
\mathcal{L}$ inside the superconductor. We then find
\begin{equation}
\begin{split}
	\mathcal{L} \ket{\varphi^{(e)}_\text{in}(x)}
	=&
	-S_{eh}^*\ket{\varphi^{(e)}_\text{in}(x)}\\
	&+
	S_{ee}^* T_{\md x/v_F}
	\mathcal{L}\ket{\varphi^{(e)}_\text{in}(x +\md x)},
 \end{split}
 \label{eq:app_phi_in}
\end{equation}
having used  Eq.~(\ref{eq:in-out_app}) to write
\begin{equation}
\mathcal{C}\ket{\varphi^{(h)}_\text{out}(x)}
	=
	\mathcal{L} \ket{\varphi^{(e)}_\text{in}(x)}.
\end{equation}
In addition, we can write the incoming state as
\begin{equation}
\begin{split}
	\ket{\varphi^{(e)}_\text{in}(x + \md x)}
	=&
	S_{ee} T_{\md x/v_F} \ket{\varphi^{(e)}_\text{in}(x)}\\
	&+ S_{eh} \mathcal{L} \ket{\varphi^{(e)}_\text{in}(x + \md x)}.
  \end{split}
\label{eq:app_phi_in_dx}
\end{equation}
We can then insert Eq.~(\ref{eq:app_phi_in_dx}) into Eq.~(\ref{eq:app_phi_in}) to obtain 
\begin{equation}
\begin{split}
	\mathcal{L} \ket{\varphi^{(e)}_\text{in}(x)}
	=&
	- S_{eh}^* \ket{\varphi^{(e)}_\text{in}(x)} \\
	&+
	|S_{ee}|^2 T_{2 \md x/v_F} \mathcal{L}
	\ket{\varphi^{(e)}_\text{in}(x)}\\
	&+
	S_{ee}^* S_{eh} T_{\md x/v_F} \mathcal{L}^2
	\ket{\varphi^{(e)}_\text{in}(x+\md x)},
  \end{split}
\label{eq:app_phi_in_2}
\end{equation}
having used that $\mathcal{L}$ commutes with $T_{\delta t}$.

Now, to lowest order in $\md x$, we have $S_{eh} \simeq \mi \Delta  \md x/\hbar v_F$ and $S_{ee} \simeq 1$. In the frequency domain we also have $T_{\md x/v_F} \simeq 1 - \mi \omega \md x/v_F$. Assuming that
$\ket{\varphi^{(e)}_\text{in}(x)}$ varies smoothly, such that $\ket{\varphi^{(e)}_\text{in}(x+\md x)}\simeq \ket{\varphi^{(e)}_\text{in}(x)}+\partial_x \ket{\varphi^{(e)}_\text{in}(x)} \md x $,  we can rewrite 
Eq.~(\ref{eq:app_phi_in_2}) as
\begin{equation}
	\frac{\Delta^*}{\hbar} \ket{\varphi^{(e)}}
	+ 2\omega \mathcal{L} \ket{\varphi^{(e)}}
	+ \frac{\Delta}{\hbar} \mathcal{L}^2 \ket{\varphi^{(e)}}
	= 0,
\end{equation}
where we have kept only terms to lowest order in $\md x $. This expression is equivalent to Eq.~(\ref{eq:NS_resp}) of the main text.

\bibliography{biblio}

\end{document}